GLEX-2021,5,2,8,x62196

# Optimisation of a Hydrodynamic SPH-FEM Model for a Bioinspired Aerial-aquatic Spacecraft on Titan

## James McKevitt[a]*

[a] *Institute of Astrophysics, University of Vienna, Türkenschanzstrasse 17, 1180 Wien, Austria*
\* Corresponding Author

## Abstract

Titan, Saturn's largest moon, supports a dense atmosphere, numerous bodies of liquid on its surface, and as a richly organic world is a primary focus for understanding the processes that support the development of life. In-situ exploration to follow that of the Huygens probe is intended in the form of the coming NASA Dragonfly mission, acting as a demonstrator for powered flight on the moon and aiming to answer some key questions about the atmosphere, surface, and potential for habitability. While a quadcopter presents one of the most ambitious outer Solar System mission profiles to date, this paper aims to present the case for an aerial vehicle also capable of in-situ liquid sampling and show some of the attempts currently being made to model the behaviour of this spacecraft.

## 1. Introduction

Saturn's largest moon Titan is unique in our Solar System through its support of a dense atmosphere and bodies of surface liquid, found nowhere else outside Earth [1]. The hugely successful Cassini-Huygens mission revealed the Saturnian System in tremendous detail, answering many questions, and posing many more. In-situ exploration of Titan, a cornerstone of this mission's success, was vital in the better understanding of the moon, and such accurate measurements of surface and atmospheric properties would not have been possible without it. The relative impenetrability of Titan to Cassini's radar also means surface landings, or at least atmospheric missions, are required to understand more about this intriguing world.

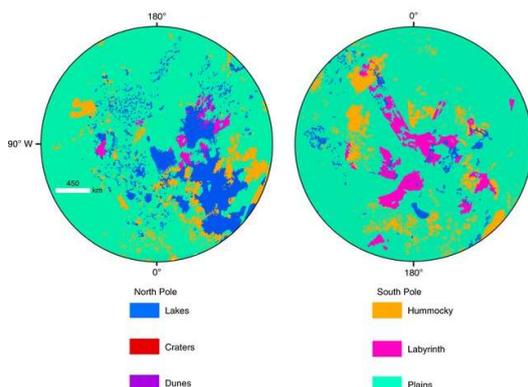

Fig. 1. Map of Titan's major geomorphological units. Adapted from Lopes et al. [2]

The equatorial regions previously benefited from a single visit by the Huygens probe and will do so again when in 2036 the Dragonfly quadcopter touches down in this area [3]. With surface liquid present in the polar regions, primarily the northern pole (Fig. 1), it will be for future missions to perform the in-situ measurement of lakes and address key science questions which will remain unanswered for a considerable period of time.

## 2. Context

ASTrAEUS (Aerial Surveyor for Titan with Aquatic Operation for Extended Usability) (Fig. 2) is a proposed mission for operation in Titan's near-surface atmosphere and bodies of surface liquid [4,5].

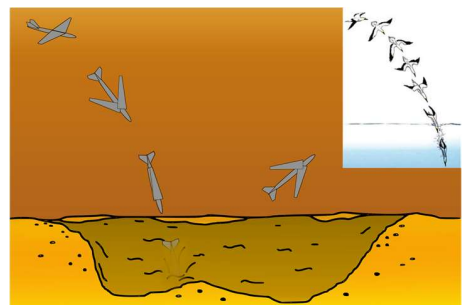

Fig. 2. Impression of a 'plunge-diving' manoeuvre by an aerial-aquatic vehicle inspired by the gannet seabird (inset) [4].

Numerous proposals have previously aimed to deliver both lake and atmospheric access, such as the joint Titan Saturn System Mission (TSSM) [6] TiME (Titan Mare Explorer) [7] proposal, or focus heavily on in-situ lake measurements, such as with the Titan Submarine [8]. ASTrAEUS, however, aims to provide a single platform capable of access to both, removing problems such as instrument duplication, and providing a powered go-to science capability in both mediums. The science goals for the mission include:

- Characterise Titan's lakes and determine their impact on the hydrological cycle
- Find the source of Titan's methane

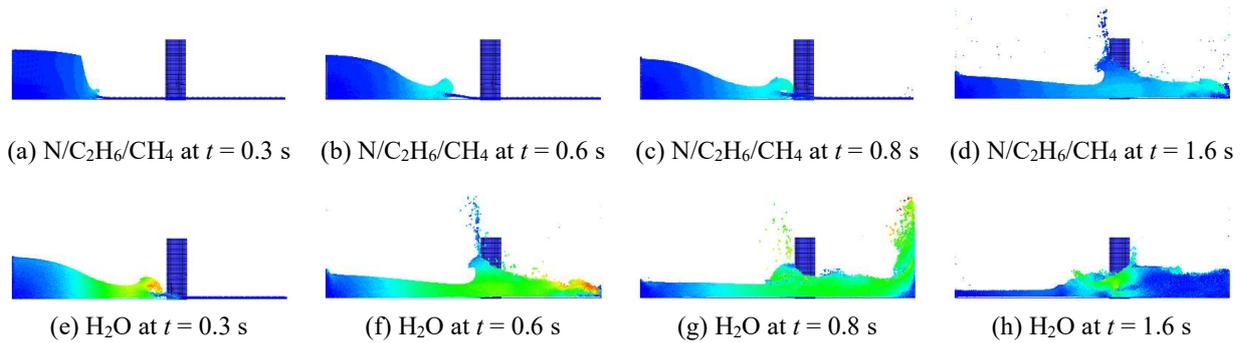

(a) N/C$_2$H$_6$/CH$_4$ at $t$ = 0.3 s  (b) N/C$_2$H$_6$/CH$_4$ at $t$ = 0.6 s  (c) N/C$_2$H$_6$/CH$_4$ at $t$ = 0.8 s  (d) N/C$_2$H$_6$/CH$_4$ at $t$ = 1.6 s

(e) H$_2$O at $t$ = 0.3 s  (f) H$_2$O at $t$ = 0.6 s  (g) H$_2$O at $t$ = 0.8 s  (h) H$_2$O at $t$ = 1.6 s

Fig. 3. Wave development and structure interaction of Titan liquid and Earth water with velocity displayed. Red corresponds to a higher relative velocity.

- Characterise Titan's near-surface atmosphere

This method is obviously not without its complications, and a plethora of problems are presented with this method. It should be noted that while the funding of Dragonfly might initially discount any near-future missions to Titan, the recent selection of two Venus missions for the NASA Discovery Program [9] does not remove the possibility entirely, although the higher cost of the former is acknowledged.

## 3. Study
### 3.1 Scope

This work presents refinements to the previous numerical models of McKevitt [4], using a coupled smoothed-particle hydrodynamics (SPH) finite element method (FEM) approach to study the 'plunge-dive' manoeuvre of ASTrAEUS on landing, well suited to the respective high surface liquid deformation and spacecraft rigidity.

Previous work, limited by project time and computing power, provided a preliminary comparison between the spacecraft's performance in water and liquid nitrogen-ethane-methane. Now, with dedicated research time and use of the Vienna Scientific Cluster, much larger and more detailed simulations more accurately replicate impact forces and fluid responses during the 'plunge-dive' landing of ASTrAEUS.

Multiphysics simulations are performed in LS-DYNA, with refinements centring on equation of state implementation and contact behaviour. An extension of accurate simulation time is also possible, thanks to a much-expanded simulation domain.

Results are validated against performance predicted by hydrodynamic equations and experimental observations, with previous numerical work using SPH techniques also considered.

### 3.2 Methodology

SPH presents an attractive method for free-surface penetration modelling given the high boundary deformation of the liquid surface. In this method, a continuum is approximated by a collection of particles interacting within a set domain and whose strength of interaction is proportional to their proximity (Fig. 4).

In this study, the behaviours of particles acting under two momentum regimes were observed, those seen in Equ. 1 and 2.

$$\frac{dv^\alpha}{dt}(x_i) = \sum_{j=1}^{N} m_j \left( \frac{\sigma^{\alpha,\beta}(x_i)}{\rho_i^2} A_{ij} - \frac{\sigma^{\alpha,\beta}(x_j)}{\rho_j^2} A_{ji} \right) \quad (1)$$

$$\frac{dv^\alpha}{dt}(x_i) = \sum_{j=1}^{N} m_j \left( \frac{\sigma^{\alpha,\beta}(x_i)}{\rho_i \rho_j} A_{ij} - \frac{\sigma^{\alpha,\beta}(x_j)}{\rho_i \rho_j} A_{ji} \right) \quad (2)$$

It is now better understood, as was previously postulated, that Equ. 2, the momentum formulation known as the 'fluid formulation', provides a more stable and efficient solution in the case of this impact problem. It is better able to cope with the high-velocity impact and subsequent shockwave propagation through the fluid, and as such provides more accurate results.

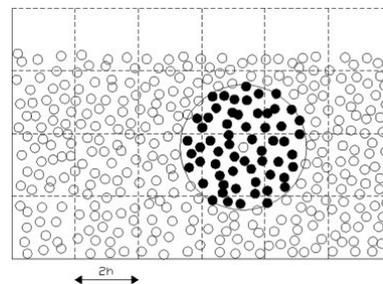

Fig. 4. SPH particles in 2D, with interacting neighbours, highlighted. Reproduced from Gomez-Gesteira et al. [10].

Additional refinement to the contact behaviour between the projectile and surface liquid has also led to a reduction in the oscillation of impact retardation and oscillation throughout the projectile itself. This is mainly due to additional 'smoothing' constraints being placed on the projectile boundary, and its specification as a shell part as opposed to a solid.

*3.3 Results*

These additional simulation constrains present no dramatic change in behaviour of the surface liquid, when compared to Earth liquid for a rudimentary comparison, as seen in Fig. 3.

Of key interest in this study is the conclusion that the spacecraft will passively descent to a depth of approximately 50m, and the smoothing of the deceleration, compared with the somewhat erratic deceleration seen previously. This can be viewed as confirmation that a more realistic physical behaviour of the projectile is replicated in current models.

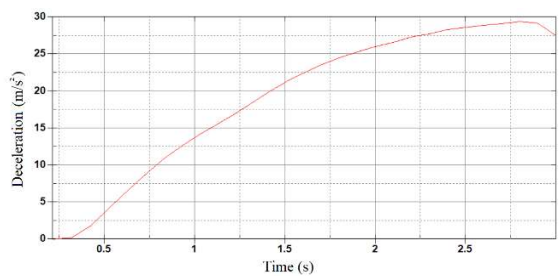

Fig. 5. Deceleration of projectile entering Titan liquid

**4. Conclusion**

Augmentations to the SPH model for Titan's surface liquid presented by McKevitt [4] have been made, with parallelization of simulations allowing for a much enhanced resolution and better understanding of previously poorly understood phenomenon. Strength of oscillations in the spacecraft fuselage have been better understood, and their amplitude is of less significant impact to the 'plunge-dive' maneuver than was first evaluated. The spacecraft can be expected to penetrate to a depth of approximately 50m. Further advocation of the mission concept should continue.

**Acknowledgements**

Thanks are extended to the Royal Astronomical Society and the Royal Academy of Engineering for their continued support of this project.

Thanks are given to the University of Vienna and its Institute of Astrophysics for their hosting of this research, and to the Star and Planetary Formation Group for their input. The Vienna Scientific Cluster is acknowledged for its use, training courses, and for providing an LS-DYNA licence.